%% file: WP.tex
\pdfoutput=1
\documentclass[a4paper, 12pt]{article} % Font size (can be 10pt, 11pt or 12pt) and paper size 

\usepackage{geometry}
\usepackage{graphicx}
\usepackage{lmodern, textcomp}
\usepackage{amssymb}
\usepackage{booktabs}
\usepackage{amsmath}
\usepackage[T1]{fontenc}
\usepackage{tikz}
\usetikzlibrary{calc,matrix}
\usepackage{caption}
\usepackage [latin1]{inputenc}
\usepackage{fancyhdr}
\usepackage{booktabs,enumitem}
\usepackage{rotfloat}
\usepackage{color}
\usepackage{subcaption}
\usepackage{eurosym}
\usepackage[authoryear]{natbib}
\usepackage[doublespacing]{setspace}
\usepackage[hyphens]{url}
\usepackage{tabularx}

\begin{document}
\title{Hiding Behind Machines: \\ When Blame Is Shifted to Artificial Agents}
\author{Till Feier\thanks{TUM School of Governance, TU Munich, Richard-Wagner-Stra\ss e 1, 80333 Munich, Germany, till.feier@tum.de} \and Jan Gogoll\thanks{Bavarian Institute for Digital Transformation / TU Munich, Gabelsbergerstr. 4, 80333 Munich, Germany, jan.gogoll@bidt.digital} \and	Matthias Uhl\thanks{ZD.B Junior Research Group ``Ethics of Digitization'', TUM School of Governance, TU Munich, Richard-Wagner-Stra\ss e 1, 80333 Munich, Germany, m.uhl@tum.de}}
\date{\today}
\maketitle

%\linespread{2}
\singlespacing

\abstract{The transfer of tasks with sometimes far-reaching moral implications to autonomous systems raises a number of ethical questions. In addition to fundamental questions about the moral agency of these systems, behavioral issues arise. This article focuses on the responsibility of agents who decide on our behalf. We investigate the empirically accessible question of whether the production of moral outcomes by an agent is systematically judged differently when the agent is artificial and not human. The results of a laboratory experiment suggest that decision-makers can actually rid themselves of guilt more easily by delegating to machines than by delegating to other people. Our results imply that the availability of artificial agents could provide stronger incentives for decision makers to delegate morally sensitive decisions.}

\vspace{0.5cm}
\expandafter\def\expandafter\quote\expandafter{\quote\small\singlespacing} %Quotes sind einzeilig

\begin{quote}

\textit{``Once the rockets are up, who cares where they come down? That's not my department'' says Wernher von Braun.}  \\    \vspace{0.2cm}
-- Tom Lehrer
 \end{quote}

\doublespacing
\newpage
\input{Introduction}\label{Intro}
\input{TheoreticalBackground}
\input{Design}

\input{Results}

\input{Conclusion}

%\input{Discussion}

%Add sections as you please - There needs to be a file called *.tex with identical name etc. in the same folder

\newpage
\bibliographystyle{chicago}
\bibliography{references}
\end{document}

%% file: Introduction.tex
\section{Introduction}

In 2017, the airline AirBerlin, Lufthansa's biggest competitor for German domestic flights, went bankrupt. Lufthansa passengers soon noticed a significant spike in ticket prices. The resulting backlash was enormous and subsequently led to an investigation by Germany's Federal Cartel Office (FCO). Lufthansa was quick to blame its automated booking system, which had allegedly responded to a spike in demand. This did little to quell the public outrage and FCO president Andreas Mundt famously said that ``companies can't hide behind algorithms'' \citep{busse2017kartellamt}. Our article tests the empirical content of this normative statement.

Algorithms have become an integral part of society and are responsible for more tasks than ever before \citep{Manyika2017future}. This is not only pushing the frontiers of technology but also challenging our traditional concepts of guilt and responsibility. The diffusion of responsibility between a multitude of actors has long been a trademark of modern institutions. The challenge of locating responsibility within a complex system may even become exponentially more difficult when moral agency is distributed between human operators and autonomous agents \citep{nissenbaum1996accountability}. The fact that even the designers of algorithms cannot fully explain their decisions  (e.g., black box models in machine learning) adds to this complexity \citep{rudin2019stop}.

Normative ethics raise a fundamental question with respect to the increasing use of artificial agents in decision-making. In which sense can artificial agents also be moral agents and therefore be responsible for their actions? Philosophers and engineers started pondering this question decades ago when computers were merely functioning as calculators \citep{moor1979there}. There is still little consensus on the matter and the idea of ``moral machines'' remains under debate \citep{allen2012moral}. In any case, the philosophical discussion is predominantly concerned with normative aspects and the question of who ought to be held responsible or blamed if a machine brings about moral evil. 

Although these normative questions are of obvious importance, the behavioral impact that the interaction with artificial agents has on the operator's conduct is not well understood either. It seems crucial, however, for research on behavioral consequences to feed back into the ethical debate. Human operators might get trapped between the increasing capabilities and autonomy of artificial agents on one hand and our rigid moral understanding of guilt and responsibility on the other. Operators may end up merely filling responsibility gaps in various systems instead of actually being in charge. They might also over-utilize artificial agents even when they are not the ideal choice for a given task, because they wish to use machines as scapegoats \citep{danaher2016robots}. 

This study investigates whether delegators will be able to successfully shift blame by delegating tasks to artificial agents. Our investigation focuses on whether people judge delegations to human and artificial agents differently in light of given outcomes. We find no differences between judgments toward human and artificial agents in the event of good outcomes. This means that the morally beneficial delegation to an artificial agent was considered neither better nor worse than the beneficial delegation to a human agent. However, if a bad outcome occurred, delegators fared significantly better if a machine agent caused the failure. Interestingly, participants did not seem to anticipate this pattern as we did not find significant differences regarding delegation decisions themselves. In fact, decisions involving human and artificial agents seem driven by the expected utility of the delegation for the affected party.

Our article proceeds as follows. In Section 2, we will derive our research question. In section 3, we will outline the experimental setup to test it. We will discuss our results in Section 4 and conclude in Section 5.

%% file: TheoreticalBackground.tex
\section{Background and Research Question}

A substantial amount of literature is available on the parameters that influence automation use in teams of human supervisors and the machines at their disposal \citep{dzindolet2002perceived}. In recent years, a number of studies investigated how those parameters change once decisions have moral undertones, i.e., an impact on third parties. \citet{goldbach2019transferring} found people to be hesitant to delegate decisions to algorithms that affect both the decision-maker herself as well as third parties. Similarly, a study by \citet{niszczota2020robo} suggests that algorithm aversion extends to the financial sector, and that people especially prefer human over artificial agents when it comes to making financial decisions with moral undertones. In a laboratory study, \citet{gogoll2018rage} identified a strong aversion against delegating morally relevant tasks to algorithms. It seems that people were less willing to delegate tasks to machines if those decisions imposed monetary externalities on third parties. While their study checked for ``perceived utility'' of the artificial agent and the trust in that agent, they were unable to determine the exact causes of the profound algorithm aversion. 

Other studies also support the idea that a lack of trust is unlikely to be the cause of aversion towards machine use. If anything, there seems to be over-reliance and over-trust in machines, even if the lives of people are at stake \citep{robinette2016overtrust}. This suggests that there have to be other causes for machine aversion in decisions with moral implications. We hypothesize that responsibility is a key concept in understanding this phenomenon. Perceived responsibility is already an important research topic in relation to machine use, probably most prominently in the context of automated driving \citep{hevelke2015responsibility}. However, little attention has been paid to the question of how the introduction of machine agents might affect the blame and praise that people ascribe to the delegator in light of a given outcome. Understanding how the use of artificial agents is judged would be an important first step in gaining a deeper understanding of how their availability might influence people's motivation to delegate morally sensible tasks. This is closely linked to the idea that avoiding blame or shirking responsibility can be pivotal factors in delegation decisions.

Strategies of blame avoidance and responsibility shifting have long been discussed in the political sciences \citep{weaver1986politics}. Instances of so-called blame games can frequently be observed in political systems with regard to policy making and implementation in the European Union \citep{heinkelmann2020multilevel} or between officials from different levels of government in the United States \citep{maestas2008shifting}. Unsurprisingly, similar strategies can also be observed in the private sector, for instance, when it comes to upholding employment standards within franchise networks \citep{hardy2019shifting}. The idea that responsibility shifting can in fact be a pivotal factor in delegation decisions plays an especially prominent role in public choice theory \citep{fiorina1986legislator}.

Experimental evidence of this phenomenon comes from \citet{fischbacher2008shifting}, who showed that responsibility attribution can sometimes be effectively shifted and that this constitutes a powerful motive for decision-makers. Other experiments provide evidence that this is true even if the agent in question is effectively powerless \citep{hill2015does}, or the delegation by the principal eliminated the possibility of a fair outcome \citep{oexl2013shifting}. But does this also hold true for artificial agents?

Popular culture places a strong emphasis on human responsibility despite an empirical decline of human control in many areas \citep{elish2015praise}. The concept of ``algorithmic outrage asymmetry'' suggests that people are less morally outraged by algorithmic wrongdoing than by human wrongdoing, for instance, in cases of discrimination by age, race or gender \citep{bigman2020algorithmic}. If, however, people seek to attribute blame, some argue that it will not be placed on algorithms but that humans could emerge as ``moral crumple zones.'' In human-machine teams, humans would then have to take on blame even for accidents outside of their control \citep{elish2019moral}. 

While some argue that humans will bear all of the responsibility and none of the control when working with machines, others think that machines are perfectly suited to be used as scapegoats. So far very limited empirical evidence has been provided to back either position. \citet{strobel2017sharing} investigated whether choices in a dictator game and perceived guilt change when players can share responsibility with machines. The authors report that perceived responsibility and guilt did not vary significantly when comparing purely human teams with human-machine teams. They write that people tend to make fewer selfish decisions when partnered with machines, though that effect was insignificant. 

So while responsibility shifting has long been established as an integral part in delegation decisions, empirical evidence of whether the introduction of artificial agents is rendering this motive mute or even more important is lacking. This constitutes a serious research gap, the implications of which exceed academic relevance. A better understanding of responsibility shifting to artificial agents could explain over and underreliance on machines and profoundly influence legal decision-making regarding automation and digitization. For instance, administrators are struggling to provide governance strategies for automated vehicles, because of the ambiguity with respect to liability \citep{taeihagh2019governing}. A better understanding of how people actually attribute responsibility is likely to help create guidelines that are not only more effective but also more likely to gain consensus.

Deeper insights into the phenomenon might also help to shield human operators from unjust recrimination. As mentioned above, our classic understanding of human-machine teams has cemented a focus on human responsibility despite a decline of human control in various areas \citep{elish2015praise}. This is especially troubling since responsibility has proven to be be an important factor in understanding and predicting punishment patterns. (for example \citep{fehr1999theory,bolton2000erc}) It seems that human operators are in fact in harms way and might become scapegoats in cases of technical failure.

In contrast, machine use might emerge as a strategy for self-exculpation in critical situations. This could have detrimental effects if it leads to an overuse of machines - a bleak prospect given the growing capabilities of algorithms and the potential harm this implies for workers and consumers. In sum, there are plenty of reasons to investigate attributions of blame and praise in the context of automation. 

To shed light on this problem, we will test the following conjecture in a laboratory experiment.\\
\\ 
\textbf{Conjecture:} Delegators are rewarded differently for delegating other-regarding tasks to artificial as compared to human agents.\\
\\
On one hand, the delegation of a task that could carry severe consequences for a third party to an unmonitored machine might be considered careless and result in punishment or defamation. On the other hand, principals might be exculpated entirely since people deem the failure of machines to be even more outside of their control than when delegating to another human. Either way, a better understanding of public reservations regarding the introduction of novel technology is likely to prove useful in future moral and legal considerations. The experiment designed to test the above conjecture will be outlined in the following chapter.

Additionally, we will explore whether the effect of perceived utility on delegation decisions varies depending on the agent's artificial or human nature. Based on a definition by \citet{dzindolet2002perceived}, we define the perceived utility of employing an artificial agent as the difference between the perceived reliability of an automated device and the perceived reliability of manual control. Furthermore, we will include risk attitudes into our analysis \citep{o2018modeling}. 

%% file: Design.tex
\section{Design}

The experiment consisted of (1) a logic task, (2) a delegation decision, (3) an evaluation of the delegation decision, (4) a self-assessment of one's performance and elicitation of risk attitudes. Subjects received a 40 ECU show-up fee increased by the outcome of the delegation (successful or not), plus (minus) their reward (their punishment) for their decision to delegate or not and the bonus for the accuracy of their self-assessment and the lottery as payment for the experiment. An experimental currency unit was used (ECU) with the exchange rate of 10 ECU for 1 EUR.

Subjects received their instructions on-screen and were fully informed about the rules of the game. They were randomly matched according to perfect stranger matching, i.e., no two participants interacted more than once during the experiment.
The experimental manipulation consisted of changing the nature of the agent to which a task could be delegated: it was either another human participant or an artificial agent (see \ref{Delegation}). 
Table \ref{tab:overview} gives an overview over the four stages of the experiment.

\begin{table}[h]
\centering
\begin{tabularx}{0.86\linewidth}{|X|X|}
\hline
\textbf{Stage} & \textbf{Human (Machine) Treatment}                                                                                        \\ \hline
(1) Solve logic task     &  Participants solve a series of logic puzzles                                                    \\ \hline
%Display of Results     & Subjects see the results of human participants (the machine) shown in a histogram                                                   \\ \hline
(2) Make delegation decision     & Participants decide whether to delegate to another human (a machine) or to have their own work count       \\ \hline
(3) Evaluate delegation decision     & Participants reward or punish decision to delegate or not                                                                  \\ \hline
(4) Self-assess \& choose lotteries     & Participants guess how many errors they made in logic task and reveal their risk attitude by choosing between lotteries \\ \hline
\end{tabularx}
\caption{Overview of experimental stages}
\label{tab:overview}
\end{table}

\subsection{Logic Task}

\begin{figure}[h]
\centering
\includegraphics[width=14.0cm]{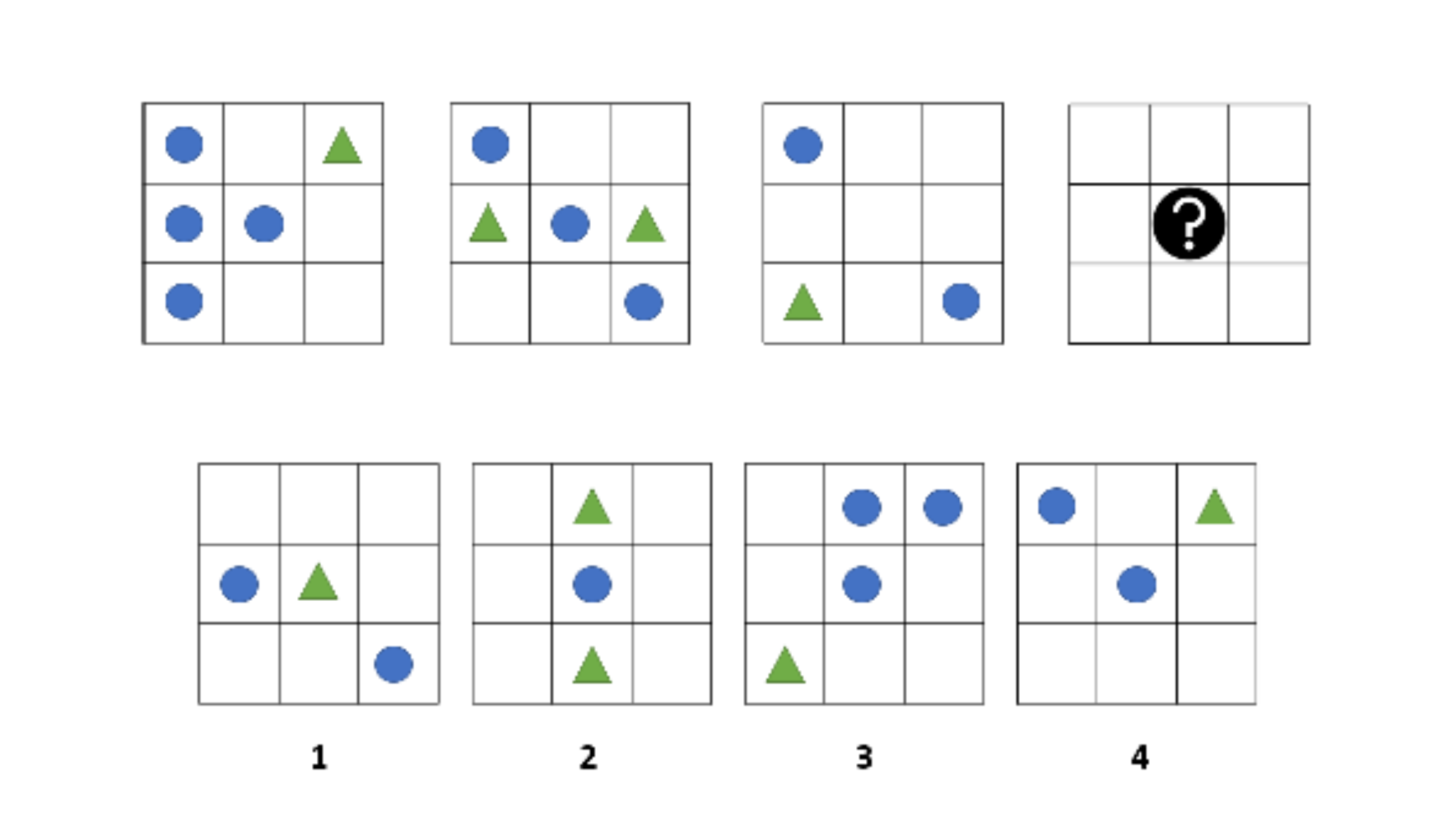}
\caption{Example of a logic task}
\label{fig:logic}
\end{figure}

The main task of the experiment was a logic puzzle. We asked participants to complete ten puzzles within a five-minute time frame. Each puzzle consisted of a sequence of three patterns and a placeholder for the missing fourth pattern. The right answer had to be derived from the given sequence and selected from a set of four alternatives. To identify the right answer, participants always had to focus on the circles, while ignoring differently shaped symbols and any colors (see Figure \ref{fig:logic}). Because we assumed that delegation decisions would heavily depend on the agent's perceived capabilities, we chose a task involving visual perception to foster participants' intuition that the algorithm could err. Participants had to complete ten of the tasks listed above before they could continue to the second stage of the experiment.

\subsection{Delegation Decision}\label{Delegation}

After the logic task had been completed, participants were randomly assigned to one of two treatments: the ``human'' or the ``machine'' treatment. Only at this point did subjects learn about the nature of the agent to whom they could delegate the task.

To give participants an idea of their potential delegatee's capability, they were shown representations of the agent's performance. In the human treatment, the $n$ participants were shown a histogram displaying the performance of the other $(n - 1)$ participants based on the actual results of the running session. In the machine treatment, participants were shown a histogram with information about how often the algorithm failed to give correct answers in $n$ trial runs. The algorithm was programmed such that it mirrored the performance of the human participants in the room. The performance of the artificial agent was therefore as good as that of the participants in the respective session. This process ensured that the delegation decisions were based on the agent's nature instead of any assumptions about differing capabilities. 

Participants were then asked to make the delegation decision. They chose whether their own performance or that of their human or artifical agent (depending on the treatment) would determine the third party's payoff. The delegation decision would thus not affect the payment of the delegator but that of another participant.
Conditional on participants' decisions to delegate or not, one of the ten solutions to the ten puzzles was randomly chosen from the answers of their agent or their own answers. If the selected puzzle was solved correctly, the third party received an additional payoff. If it was solved incorrectly, the third party did not receive an additional payoff.\vspace{0.8cm}

\begin{figure}[h]
\centering
\includegraphics[width=14.0cm]{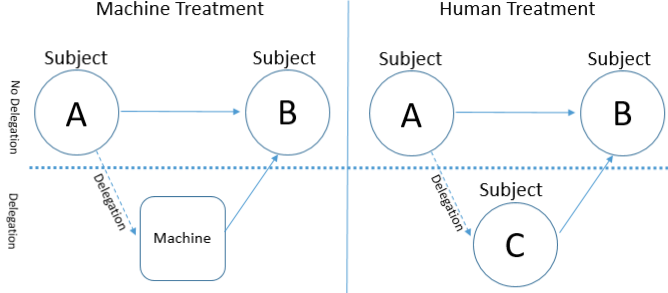}
\caption{Delegation decision for human and machine treatment}
\label{fig:del}
\end{figure}

\subsection{Evaluation of the Delegation}\label{section:reward}

Participants were now given the opportunity to punish or reward another participant's decision to delegate or not. Participants did not know the actual decision of the participant that they evaluated, nor whether this decision had resulted in a good or bad outcome for themselves. They were asked to alter the payment of the participant upwards or downwards by at most 40 ECU. Reward and punishment choices were contingent on the two possible decisions to delegate or not and the two possible outcomes of success or failure. Thus, in any case, four choices had to be made. Only the one that applied to the actual decision of the evaluated participant combined with actual outcome did finally apply \citep{selten1967}. Table \ref{tab:my-table} depicts the table that subjects saw on their screen.\vspace{0.2cm}

\begin{table}[H]
\centering
\begin{tabular}{|l |c|c|}
\hline
 \textbf{Lower or increase subjects payoff given that:} & \textbf{Amount} \\ \hline
 Subject used own work -- outcome: success & *enter amount*   \\ 
 Subject used own work -- outcome: failure & *enter amount*   \\ 
 Subject delegated (machine/human) -- outcome: success & *enter amount*   \\ 
  Subject delegated (machine/human) -- outcome: failure & *enter amount*  \\ \hline
\end{tabular}
\caption{Decision to increase or decrease the delegator's pay-off}
\label{tab:my-table}
\end{table}

\subsection{Self-Assessment and Risk Attitudes}

Subsequently, participants were asked to assess their own performance by estimating how many mistakes they had made during the logic task in the experiment's first part. This estimate was incentivized by an additional payment of 50 ECU if they guess correctly. This procedure allowed us to analyze the effect of self-assessments on delegation decisions. 
As is standard in incentivized economic experiments, the experiment was concluded by an elicitation of participants' risk attitudes.\footnote{We used the procedure introduced by  \citet{holt2002risk}. The task is based on ten choices between pairs of lotteries. The potentional payoffs for the safe lottery range from \euro2.5 to \euro1.6 and are therefore always less extreme than those for the risky lottery that range from \euro4.35 to \euro0.10. For the first pair of lotteries, the probability of the high payoff is equally low in both lotteries but equally increases for both lotteries with each new pair. A participant should switch to the risky lottery once the probability for the high payoff is sufficiently high according to his or her personal risk attitude. The later a participant switches to the risky lottery, the more risk averse he or she is.}

%% file: Results.tex
\section{Results}

The experiment took place in a major German university between February and May 2019. A total of 149 subjects participated in six sessions, 43\% were female and the average age was 23.08 years (sd = 3.83). Participants received a show-up fee of \euro 4.00 and could earn additional money in the experiment. A session lasted about 45 minutes and the average payment was about \euro 13.50  per participant. The experiment was programmed in z-Tree \citep{fischbacher2007} and subjects were recruited via ORSEE \citep{greiner2004}. Data analysis was conducted using Python's numpy, scipy, and statsmodels.api libraries. The preprocessed data set and the code are available online.\footnote{https://doi.org/10.5281/zenodo.4446581}

Remember that after the delegation decision, all participants were asked to evaluate the decision of the participant who was responsible for their own pay-off. Responsibility arose from the decision to either delegate (to a human or machine) or rely on one's own work (in both treatments). Notice again that participants were informed whether they were randomly assigned to the human or machine treatment, but not whether the other participant had actually delegated or not. They were therefore asked to judge the decision with respect to the four possible outcomes according to the so-called strategy method (see Section \ref{section:reward}). This means that subjects could increase (or deduct) the payment of their responsible participant by an integer between 0 and 40 for the cases of her (1) having brought about a good outcome herself or (2) having brought about a bad outcome herself. Furthermore, they were increasing (or deducting) the payment for the case of her (3) having delegated to an agent causing a good outcome on her behalf or (4) having delegated to an agent causing a bad outcome on her behalf.

To test our conjecture that delegators are rewarded differently for delegating other-regarding tasks to artificial as opposed to human agents, we contrast evaluations between both treatments.

 Let us first consider the good-outcome case. In the human treatment, participants that did not delegate and caused the good outcome themselves were, on average, rewarded 20.28 ECU (sd = 20.85). Participants that delegated to another human who then brought about the good outcome on their behalf were rewarded 20.94 ECU (sd = 21.34). This difference is insignificant (p = 0.720, paired t-test).

 In the machine treatment, participants that did not delegate and caused the good outcome themselves were on average rewarded 23.78 ECU (sd = 20.14). Participants that delegated to a machine were rewarded 27.00 ECU (sd = 16.71). This difference is again insignificant (p = 0.106, paired t-test). \\
 \\
 \textbf{Result 1:} Delegators did not lose any recognition for a good outcome if it was caused by either their human or their artificial agent.\\
 \\
 Let us now consider the bad-outcome case. In the human treatment, participants that did not delegate and caused the bad outcome themselves were on average rewarded 7.29 ECU (sd = 22.59 ECU). Participants that delegated to another human were on average rewarded with 8.26 ECU (sd = 21.56 ECU). This difference is insignificant (p = 0.559, paired t-test).
 
 In the machine treatment, participants that did not delegate and caused the bad outcome themselves were on average rewarded 8.53 ECU (sd = 26.52). Participants that delegated to a machine were on average rewarded 12.96 ECU (sd = 24.44). This difference is significant (p = 0.041, paired t-test).
 
\begin{figure}[]
\centering
\includegraphics[width=14.0cm]{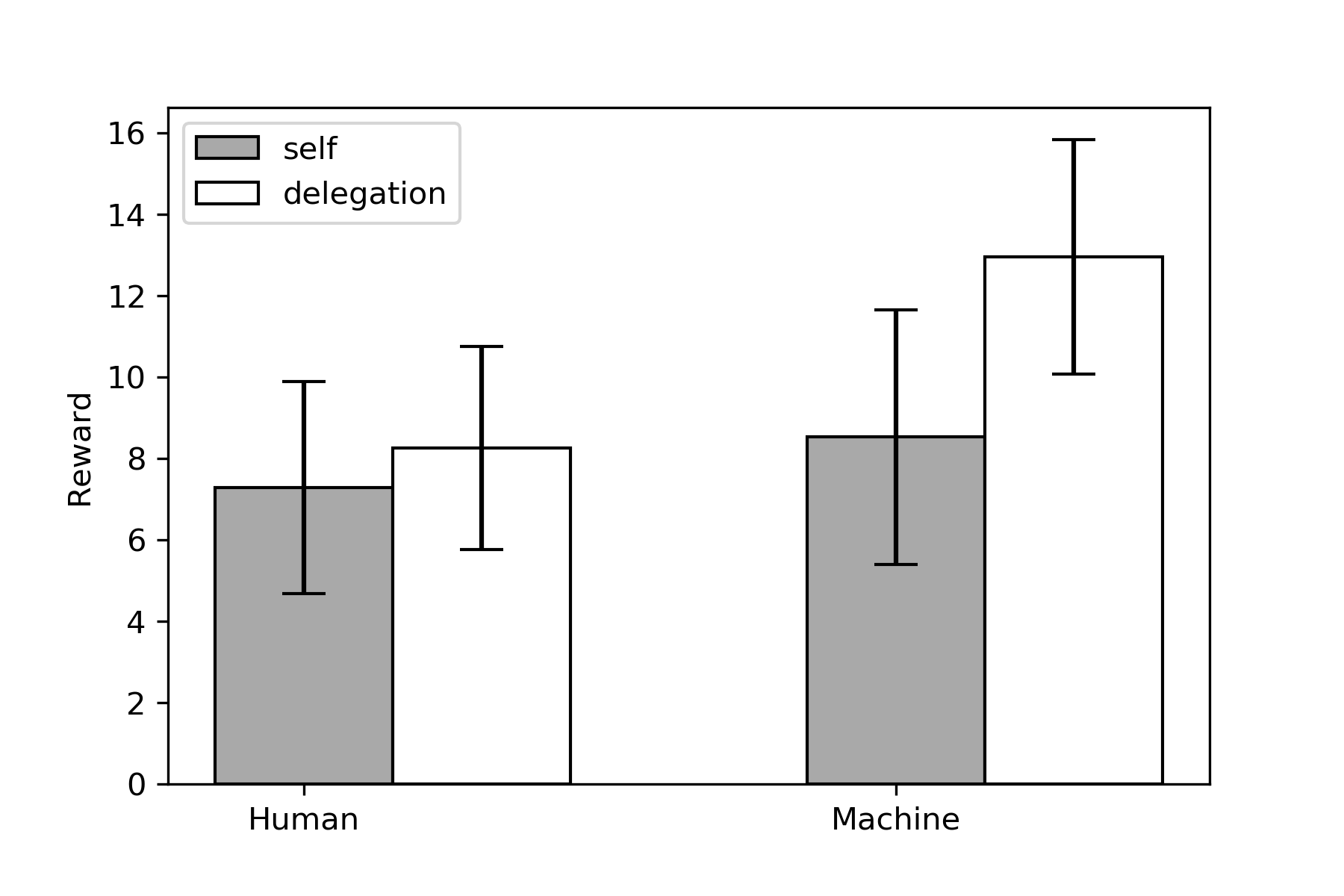}
\caption{Reward decisions if the bad outcome prevails.}
\label{fig:reward}
\end{figure}

 In the machine treatment, delegators earned higher rewards if a machine agent caused the failure. This confirms our conjecture stated in Section 2. Principals are rewarded differently for delegating tasks depending on the nature of the agent. More specifically, they fare better if a machine agent caused a bad outcome than if they did so themselves.\\
\\
\textbf{Result 2:} Delegators did not successfully shift blame for a bad outcome if their human agent caused it instead of themselves. They did, however, successfully shift blame for a bad outcome if their artificial agent caused it instead of themselves.\\
\\
Figure \ref{fig:reward} illustrates this asymmetry in termsn of rewards between the human and the machine treatment for the bad-outcome case.
 
To see whether participants exploited this incentive to delegate to machine agents that does not exist for human agents, we compare the proportions of delegators in both treatments. In the human treatment, 35 out of 76 (46.1\%) delegated the task to another human participant, whereas 41 out of of 73 (56.2\%) delegated to the machine agent in the machine treatment. The difference in the proportions of delegators is insignificant (p = 0.28?, Chi-Square Test for Independence). The similar proportions of delegators in both treatments suggests that the decision to delegate to the agent is not primarily driven by strategic concerns of avoiding blame in the event of a bad outcome. 

This is corroborated by the logistic regression reported in Table \ref{tab:reg}. The propensity to delegate (1 = yes) is positively predicted by participants' self-assessment of the number of errors they believe they have committed in the logic task (p < 0.001). ``Machine'' captures whether the agent is human (0) or artificial (1), which does not significantly influence this result. The effect to which self-assessment predicts the delegation decision is robust to controlling for risk attitude, although participants who are more risk averse are also more likely to delegate (p < 0.017).

\vspace{0.2cm}
\input{myreg}
\vspace{0.2cm}

It thus appears that the expected utility of delegating was the driving factor for participants' decisions to do so or not regardless of the nature of their agent. Those who were less confident regarding their own performance were more likely to delegate the task. Their estimated number of errors, which they stated in an incentivized self-assessment, had a significant impact on their delegation decision. 

As Table \ref{tab:reg_lin} shows, the self-assessment of participants is also predicted by the actual number of a subject's errors in the logic task. Whether they have a human or an artificial agent at their avail did not influence their self-assessment. Also, subjects' risk attitudes have no impact on their self assessment. These findings indicate that participants have a realistic impression about their own performance. Subjects who performed better in the logic task were accordingly less likely to delegate.

\vspace{0.2cm}
\input{myreg_lin_neu}

%% file: myreg.tex
\begin{table}
\begin{center}
\begin{tabular}{lclc}

\toprule
\textbf{Dep. Variable:}   &       Del        & \textbf{  No. Observations:  } &      149    \\
\textbf{Model:}           &       GLM        & \textbf{  Df Residuals:      } &      145    \\
\textbf{Model Family:}    &     Binomial     & \textbf{  Df Model:          } &        3    \\

\bottomrule
\end{tabular}
\begin{tabular}{lcccccc}
                          & \textbf{coef} & \textbf{std err} & \textbf{z} & \textbf{P$>$$|$z$|$} & \textbf{[0.025} & \textbf{0.975]}  \\
\midrule
\textbf{Intercept}        &      -3.2254  &        0.795     &    -4.055  &         0.000        &       -4.784    &       -1.666     \\
\textbf{Self\_Assessment} &       0.4092  &        0.105     &     3.906  &         0.000        &        0.204    &        0.615     \\
\textbf{Machine}          &       0.3228  &        0.365     &     0.885  &         0.376        &       -0.392    &        1.038     \\
\textbf{Risk}             &       0.1806  &        0.076     &     2.383  &         0.017        &        0.032    &        0.329     \\
\bottomrule

\end{tabular}
\caption{Generalized Linear Model Regression Results}
\label{tab:reg}
\end{center}
\end{table}

%% file: myreg_lin_neu.tex
\begin{table}
\begin{center}
\begin{tabular}{lclc}
\toprule
%\textbf{Dep. Variable:}    & Self_Assessment  & \textbf{  R-squared:         } &     0.269   \\
\textbf{Dep. Variable:}    & Self\_Assessment  & \textbf{  R-squared:         } &     0.269   \\
\textbf{Model:}            &       OLS        & \textbf{  Adj. R-squared:    } &     0.253   \\
\textbf{Method:}           &  Least Squares   & \textbf{  F-statistic:       } &     17.74   \\
\textbf{No. Observations:} &         149      &              &      \\

\bottomrule
\end{tabular}
\begin{tabular}{lcccccc}
                   & \textbf{coef} & \textbf{std err} & \textbf{t} & \textbf{P$>$$|$t$|$} & \textbf{[0.025} & \textbf{0.975]}  \\
\midrule
\textbf{Intercept} &       3.2012  &        0.474     &     6.749  &         0.000        &        2.264    &        4.139     \\
\textbf{Machine}   &       0.4598  &        0.276     &     1.666  &         0.098        &       -0.086    &        1.005     \\
\textbf{Error}     &       0.3955  &        0.059     &     6.741  &         0.000        &        0.280    &        0.512     \\
\textbf{Risk}      &      -0.0298  &        0.055     &    -0.539  &         0.591        &       -0.139    &        0.079     \\
\bottomrule
\end{tabular}
\caption{Linear Regression -- Influences on Self-assessment}
\label{tab:reg_lin}
%\caption{OLS Regression Results}
\end{center}

\end{table}

%% file: Conclusion.tex
\section{Conclusion}

Our findings indicate that delegators may be judged with more leniency in the event of a bad outcome if they delegate tasks to artificial agents instead of human agents. It does therefore stand to reason that machine agents can be successfully used to avoid ostracism. The FCO president's statement quoted at the beginning, i.e. that companies cannot hide behind algorithms, might not hold true empirically. Companies might well capitalize the effective shift of responsibility to algorithms if they fail. This is all the more true if they do not suffer any comparable loss of prestige as a result of the delegation in the event of success, as our data also suggest. The fact that the delegator in a between-subjects design receives a discharge if the agent is artificial but not if the agent is human at least suggests that the corresponding judgment is not reflective, but that it is a subtle behavioral tendency. If this is true, the discharge is not ethically desired by the evaluator.

Our results might indicate the importance of institutional solutions that hold companies and ultimately individuals liable for the moral wrongs that their artificial agents bring about. These institutions grow even more important if they have to compensate for consumers' behavioral reluctance to attribute blame in such cases. A deeper inquiry into the reluctance to punish that we observe seems warranted by the idea that extrinsic social motivation is an important factor in moral decision-making \citep{cappelen2017face}. The ability to use artificial agents as a smoke screen might encourage various decision makers to engage in more activities that are considered undesirable by stakeholders. The reassuring fact that delegators in this experiment did not exploit the effective release from blame does not imply that others will not-especially if they see through this behavioral tendency.  

Our experiment is subject to limitations. One is its rather explorative nature, which is owed to the lack of a well-established theoretical framework: Scientists have  only recently started to address the influence of algorithms on human decision-making empirically. To the best of our knowledge, the projects by \citet{gogoll2018rage} and \citet{strobel2017sharing} cited above are the only experiments dedicated to the issue so far. Furthermore, the stylized setting used in our experiment limits the scope of the conclusion that we can draw. This concern is best explained in relation to internal and external validity. Internal validity means that the experiment is designed in such a way that it warrants conclusions about the behavior of its participants inside of the laboratory, for instance, by keeping relevant factors constant (ceteris paribus) or omitting irrelevant influences (ceteris absentibus). Experiments are externally valid if their design produces findings that are informative about behavior outside of the laboratory \citep{guala2002scope}. Some authors posit an inverse relationship between internal and external validity \citep{guala2005methodology, loewenstein1999experimental}. Our findings provide a first indication that the nature of the agent has an effect on the reward and punishment that the delegator receives. The phenomenon we observed would have to be replicated in other contexts inside and outside of the laboratory to eliminate the possibility of it being a mere artifact.

We believe that our findings represent an early step in understanding delegation decisions in a domain that is gaining relevance, i.e., the use of artificial agents in moral decision-making. More generally, our experimental results illustrate the necessity of investigating the interaction between humans and machines in behavioral settings. It is insufficient to rely on ethicists' armchair arguments and on surveys that study laymen's intuitions regarding the ethical implications of algorithms, because ethically relevant implications may arise as unintended results from the interactions between people and machines. The emerging phenomena might then be difficult to anticipate and sometimes even counter-intuitive. Further research is urgently needed to create a more conclusive picture of the subtle factors that drive our behavior when cooperating (and competing) with machines.